1# Networked Model Predictive Control Using a Wavelet Neural Network

H. Khodabandehlou, *Student Member, IEEE*, M. Sami Fadali, *Senior Member, IEEE**Abstract*— In this study we use a wavelet neural network with a feedforward component and a model predictive controller for online nonlinear system identification over a communication network. The wavelet neural network (WNN) performs the online identification of the nonlinear system. The model predictive controller (MPC) uses the model to predict the future outputs of the system over an extended prediction horizon and calculates the optimal future inputs by minimizing a controller cost function. Lyapunov theory is used to prove the stability of the MPC. We apply the methodology to the online identification and control of an unmanned autonomous vehicle. Simulation results show that the MPC with extended prediction horizon can effectively control the system in the presence of fixed or random network delay.

*Index Terms*— Networked control, Model predictive control, Wavelet networks, System identification, Lyapunov.## I. INTRODUCTION

Networked control systems are control systems in which the controller, actuator and sensor are connected through a communication network. The shared network connection between different components of the control loop yields a flexible architecture and reduced installation and maintenance costs [1]. The theory of networked control systems combines control system theory and communication theory [1].

Time delay is a salient feature of any digital control system. This delay can be due to either plant delay or computational delay [2]. The computational delay can adversely affect controller performance or cause closed loop instability [3].

The control and stability of time-delayed systems has been widely studied [4], and various control and optimization algorithms have been proposed to provide satisfactory stable performance [5]. Astrom and Wittenmark studied effects of computational delay on digital controller design [6].

Although recent advances in digital processors have mitigated the effects of computational delay, network transport delay must still be considered in the design of networked control systems [2], [7], [8]. Delay switching based methods

H. Khodabandehlou is graduate student at University of Nevada, Reno, Reno, NV, 89557, USA (e-mail: hkhodabandehlou@nevada.unr.edu)
M. Sami Fadali is professor at University of Nevada, Reno, Reno, NV, 89557, USA (e-mail: fadali@unr.edu)and parameter uncertainty based methods are two other alternatives to deal with the network induced delays [9]. The approach of [10] uses parameter uncertainty based to deal with network-induced delay and uses linear matrix inequalities (LMI) to prove the existence of a stable state feedback controller. Wang and Yang [11] model the delay as a Markov chain and model the control loop as a Markov jump system then stabilize the closed loop system using an output feedback.

Although switching based methods for time delay systems are less conservative, they are more computationally costly and are difficult to implement [11]. A combination of switching and parameter uncertainty approaches is used in [11] to avoid the computational complexity of switching based approach and the conservativeness of parameter uncertainty based approach.

Traditional digital control uses uniform sampling of the measurements over time. Although this makes analysis easy, it is not optimal in terms of network traffic [13], [14]. Astrom and Bernhardsson proposed an event triggered sampling scheme to decrease the network traffic by reducing the number of packets sent over the network [13]. The main idea of event triggered base systems is to obtain a new measurement when the closed loop system does not satisfy desired performance criteria [15]. Researchers have proposed different sampling approaches such as deadband sampling [16], self-triggered sampling [17] and error energy sampling [18] to optimize the network resource usage.

Heemels and Donkers used periodic event triggered control for a piecewise linear system and for an impulsive system, and analyzed the stability of the controller for both systems [19]. Wang et. al. proved that a network control system with L1 adaptive controller and event trigger sampling scheme can be arbitrarily close to a desired stable reference system under certain conditions [20]. Peng and Hong designed $H_\infty$ controller with non-uniform sampling period. They sampled the states of the system nonuniformly, modeled the networked control system as a time-delay system, and proved the ultimate boundedness of their controller [21]. Another method of designing $H_\infty$ controllers based on Markovian modeling of sensor and actuator was presented in [22]. An observer based $H_\infty$ controller for continuous time networked control system was presented in [23]. A new model for continuous time networked control system was introduced and the observer based controller was designed based on a new Lyapunov functional. A method of designing an $L_2$ controller for decentralized event-triggered control system was presented in [24].

Wang et. al. designed an event triggered model predictive controller for wireless networked control [25]. They derived trigger conditions and proved the stability of their controller by choosing the objective function of MPC as a Lyapunov function. A networked predictive controller to dampen power system inter-area oscillations was presented in [26]. The network predictive controller uses a generalized predictive control scheme to calculate the optimal control input for constant and random network delay. The stability analysis of a networked control system with a predictive-observer based controller was presented in [27]. They proved stability using two different Lyapunov functions, a function derived from network conditions, and a common quadratic Lyapunov function.

Cao et al. used a Gaussian process model of the unknown dynamics of a quadrotor with model predictive control [28]. Their methodology handles the model uncertainty and is computationally efficient. A locally weighted learning model predictive control (LWL-MPC) is presented in [29]. The model can effectively learn nonlinear and time varying dynamics online.

In this study, we use a wavelet neural network with feedforward component for online nonlinear system identification. Zhang and Benveniste argued that wavelet neural networks may have fewer nodes that other artificial neural networks [30]. The feedforward component drastically reduces the number of hidden layer nodes and consequently reduces the training time of the wavelet neural network. The improved computational efficiency of the wavelet networks with feedforward component makes it ideal for online identification and control applications [31]. The model predictive controller uses wavelet neural network to predict the future outputs of the system over an extended prediction horizon and minimizes a cost function to find the optimal control action. Lyapunov stability theory is used to prove the stability of the model predictive controller.

To demonstrate the efficacy of our networked control approach, we apply it to the control of an unmanned autonomous vehicle. Two scenarios of fixed and random network delay are simulated. Simulation results show that the model predictive controller with extended prediction horizon can successfully mitigate the effect of fixed and random network delay. A preliminary version of this work was presented in [32].

The remainder of the paper is organized as follows: Section II presents the networked control system. Section III describes model predictive controller and Section IV presents simulation results. The conclusion is in Section V. Appendix I provides a stability proof for the model predictive controller.

## II. NETWORKED CONTROL SYSTEM

A block diagram of networked control of an autonomous vehicle is shown in Fig. 1. The measurements are sent over the communication channel to the WNN and MPC. Measurements are received after a network delay and they may get lost due to packet loss in the communication channel. The WNN weights are updated after receiving the measurement, then the MPC predicts the future outputs of the system. Using the predicted outputs and the desired future outputs of the system, the controller calculates the future optimal control inputs by minimizing a controller cost function.

Due to network delay, the system does not promptly receive the control input. Hence the controller needs to make more predictions to compensate for the effect of delay from sensor to controller and controller to actuators.

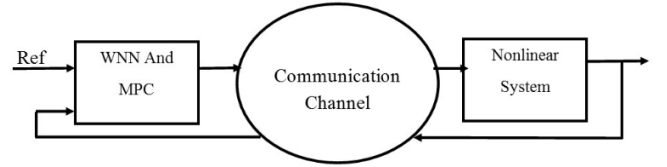

Fig. 1. General scheme of plant and controller

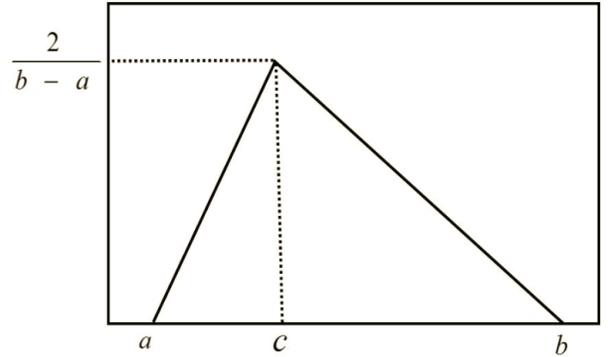

Fig. 2. Probability density function of triangular distribution

We use a feedforward wavelet neural network with one hidden layer and feedforward component to identify the model of nonlinear system. The wavelet neural network structure is shown in Fig. 3. The input-output equation of the WNN is described as

$$\hat{\boldsymbol{y}}(k) = S\boldsymbol{\psi}(\boldsymbol{u}_N) + Q\boldsymbol{u}_N \qquad (1)$$

where $\boldsymbol{u}_N = \left[u_{N_1}, \dots, u_{N_n}\right]^T$ is the input vector to the network and $\hat{\boldsymbol{y}}(k) = \left[\hat{y}_1(k), \hat{y}_2(k), \dots, \hat{y}_J(k)\right]^T$ is networks output. $n$ is number of inputs to the network and $J$ is number of network outputs. Activation function of hidden layer nodes are assumed to be Mexican hat wavelet

$$\psi_i(t_i) = \frac{2\pi^{\frac{1}{4}}}{\sqrt{3}}(1 - t_i^2)e^{-\frac{t_i^2}{2}} \qquad (2)$$

$$t_i = \frac{\boldsymbol{w}_i^T \boldsymbol{u}_N - b_i}{a_{i,i}}, i = 1, \dots, m \qquad (3)$$

where $m$ is number of hidden layer nodes. With $m$ nodes in the hidden layer, $S$, $W$ and $Q$ will be $J \times m$, $m \times n$ and $J \times n$ matrices respectively. $\boldsymbol{b} = [b_i]_{m \times 1}$ is vector of shift parameter of wavelets and $A = diag([a_{1,1}, \dots, a_{m,m}])$ is diagonal matrix whose diagonal elements are scale parameter of hidden layer activation functions.

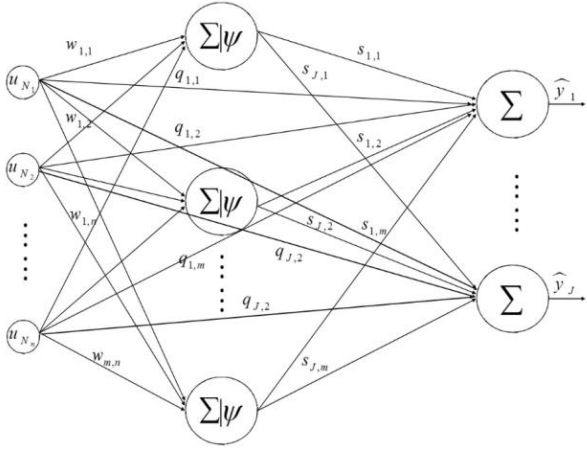

Fig. 3. Structure of wavelet neural network [32]

The feedforward component weights is tuned using well known recursive least squares algorithm and the rest of the parameters are tuned using error backpropagation algorithm. Due to fast convergence of recursive least squares algorithm no pre tuning is needed for wavelet neural network. This makes the network more interesting for online identification and control applications.

The cost function for training wavelet neural network is the sum of squared errors (SSE)

$$J_N = \frac{1}{2}\sum_{k=1}^{N} \|\hat{\boldsymbol{y}}(k) - \boldsymbol{y}(k)\|^2 \qquad (4)$$

where $\boldsymbol{y}(k)$ is the vector of measured outputs and $\hat{\boldsymbol{y}}(k)$ is the output of the WNN. Network parameters are updated by the gradient descent algorithm

$$\sigma_{t+1} = \sigma_t - \gamma \nabla_\sigma(J_N) \qquad (5)$$

$\gamma$ is the learning rate for gradient descent algorithm with $\gamma \in (0,1]$. After training the feedforward component using the recursive least squares algorithm, the network modeling error can be calculated as

$$\boldsymbol{e}_y(k) = \hat{\boldsymbol{y}}(k) - \boldsymbol{y}(k) - Q\boldsymbol{u}_N \qquad (6)$$

$$\boldsymbol{e}_y(k) = [\boldsymbol{e}_{y_1}, \dots, \boldsymbol{e}_{y_J}]^T \qquad (7)$$

The cost function for training the neural network can be rewritten in terms of the error as

$$J_N = \frac{1}{2}\sum_{k=1}^{N} \|\boldsymbol{e}_y(k)\|^2 \qquad (8)$$

The chain rule is used to calculate the gradient of the cost function with respect to the WNN parameters

$$\nabla_\sigma(J_N) = \boldsymbol{e}_y^T(k)\partial\hat{\boldsymbol{y}}(k)/\partial\sigma \qquad (9)$$

The following proposition assures the stability of the WNN in terms of the WNN learning rate.

**Proposition [32]**: A WNN is stable if the learning rate satisfies $\gamma < \frac{2}{k}$ where $k = \max_j \left\|\frac{\partial \hat{y}_j(n)}{\partial \boldsymbol{o}_j(n)}\right\|_2^2$, $j = 1,2,\dots,J$, $\hat{y}_t(n)$ is the $j^{th}$ network output at time step $n$, and $\boldsymbol{o}_j = [a_i, b_i, w_{i,j}, q_{i,j}, s_{i,j}]$ is the vector of network parameters that affect the $j^{th}$ output.

### III. MODEL PREDICTIVE CONTROLLER

To design a model predictive controller, an accurate model of the system is needed. We assume that outputs of the system can be measured in each iteration to update the feedforward component and WNN weights. The updated WNN is used to predict the future outputs of the system and the MPC uses the predicted outputs to calculate future inputs by minimizing the controller cost function. The cost function is not subject to input or state constraints because the prediction horizon, the control horizon, $\xi_j$ and $\rho_i$ provide enough flexibility to minimize the tracking error while restricting the magnitude of the control action.

The controller cost function is given by

$$J_c = \frac{1}{2}\sum_{i=1}^{N_p}\sum_{j=1}^{J} \xi_j e_{yj,c}(n+i)^2 \\ + \frac{1}{2}\sum_{j=1}^{n_u}\sum_{i=1}^{N_u} \rho_i \Delta u_j(n+i-1)^2 \qquad (10)$$

where $\boldsymbol{e}_{yj,c}$ is the prediction error defined as error between predicted outputs, $\hat{\boldsymbol{y}}_j(n+1)$, and desired outputs $\boldsymbol{y}_{j_d}(n+1)$. $\Delta u_j(n), j = 1,\dots,n_u$ is the increment in the $j^{th}$ input. $\xi_j$ and $\rho_i$ are penalty factors on the $j^{th}$ tracking error and the $i^{th}$ input change respectively. Both factors are assumed to be in range $(0,1]$. $J$ is the number of outputs and $n_u$ is the number of inputs. Small values of $\rho_i$ lead to smaller and smoother control action while large values lead to faster tracking but may cause controller instability. Small values of $\xi_j$ lead to smaller tracking error and larger control action while larger values of $\xi_j$ increase the tracking error while reducing the magnitude of the control input. $N_p$ is the prediction horizon and $N_u$ is the control horizon. Large values of the prediction horizon lead to smoother control action but increase the tracking error while smaller values lead to better tracking and larger control input.

Defining the $H$ matrix as

$$H = \begin{bmatrix} 1 & 0 & 0 & \dots & 0 \\ -1 & 1 & 0 & \dots & 0 \\ 0 & -1 & 1 & \dots & 0 \\ 0 & \dots & \ddots & \ddots & 0 \\ 0 & \dots & 0 & -1 & 1 \end{bmatrix} \qquad (11)$$

the controller cost function can be rewritten as

$$J_c = \frac{1}{2}\left[\sum_{j=1}^{J} \xi_j \boldsymbol{e}_{yj,c}^T(n+1)\boldsymbol{e}_{yj,c}(n+1) \\ + \sum_{i=1}^{n_u} \rho_i (H\boldsymbol{u}_i(n))^T(H\boldsymbol{u}_i(n))\right] \qquad (12)$$

where

$$\boldsymbol{e}_{yj,c}(n+1) = \boldsymbol{y}_{j_d}(n+1) - \hat{\boldsymbol{y}}_j(n+1), \\ 1 \leq j \leq J \qquad (13)$$

$$\mathbf{y}_{j_d}(n+1) = [y_{jd}(n+1),\ldots,y_{jd}(n+N_p)]^T, \quad 1 \leq j \leq J \tag{14}$$

$$\hat{\mathbf{y}}_j(n+1) = [\hat{y}_j(n+1),\ldots,\hat{y}_j(n+N_p)]^T, \quad 1 \leq j \leq J \tag{15}$$

$$\mathbf{u}_i(n) = [u_i(n), u_i(n+1),\ldots,u_i(n+N_u-1)]^T, \quad 1 \leq i \leq m \tag{16}$$

$\mathbf{u}_i(n)$ is updated by minimizing $J_c$ using the gradient descent algorithm

$$\frac{\partial J_c}{\partial \mathbf{u}_i(n)} = -\sum_{j=1}^{J} \xi_j G_{y_j,\mathbf{u}_i}^T(n) \mathbf{e}_{yj,c}(n+1) + \rho_i H^T \Delta \mathbf{u}_i(n), 1 \leq i \leq n_u \tag{17}$$

$$\Delta \mathbf{u}_i = \eta_i \left(-\frac{\partial J_c}{\partial \mathbf{u}_i(n)}\right)$$
$$= -\eta_i \left(I_{N_u * N_u} + \xi_i \rho_i H^T\right)^{-1} \sum_{j=1}^{J} \xi_j G_{y_j,\mathbf{u}_i}^T(n) \mathbf{e}_{yj,c}(n+1), 1 \leq i \leq n_u \tag{18}$$

$$\mathbf{u}_i(n+1) = \mathbf{u}_i(n) + \Delta \mathbf{u}_i, 1 \leq i \leq n_u \tag{19}$$

The learning rate for the gradient descent algorithm, $\eta_i$, is assumed in range (0,1]. $G_{yj,ui}$ is defined as

$$G_{y_j,u_i}(n) = [g_{y_j,u_i}(k,l)]$$
$$g_{y_j,u_i}(k,l) = \begin{cases} \frac{\partial \hat{y}_j(n+k)}{\partial u_i(n+l-1)}, & k \geq l \\ 0, & k < l \end{cases} \tag{20}$$
$$1 \leq i \leq n_u, 1 \leq j \leq J, 1 \leq k \leq N_p, 1 \leq l \leq N_u$$

The derivatives in $g_{y_j,u_i}(k,l)$ are calculated using the chain rule

$$\frac{\partial \hat{y}_j(n+q)}{\partial u_i(n+r)} = \frac{\partial \hat{y}_j(n+q)}{\partial \hat{y}_j(n+q-1)} \times \frac{\partial \hat{y}_j(n+q-1)}{\partial \hat{y}_j(n+q-2)} \times \cdots \times \frac{\partial \hat{y}_j(n+r+2)}{\partial \hat{y}_j(n+r+1)} \times \frac{\partial \hat{y}_j(n+r+1)}{\partial u_i(n+r)} \tag{21}$$

Appendix I provides the stability proof for the MPC.

## IV. SIMULATION RESULTS

A simple model of an unmanned autonomous vehicle is presented in Fig.4. The system has two control inputs, the speed of autonomous vehicle $v(n)$ and the steering angle $\alpha(k)$. The dynamics of the autonomous vehicle is completely controllable through the two control inputs. The input-output equation of the vehicle is described by

$$\mathbf{x}(n+1) = \begin{bmatrix} x(n+1) \\ y(n+1) \\ \theta(n+1) \end{bmatrix}$$
$$= \begin{bmatrix} x(n) + Tv(n)\cos(\theta(n)\cos(\alpha(n))) \\ y(n) + Tv(n)\sin(\theta(n)\cos(\alpha(n))) \\ \theta(n) + Tv(n)\sin(\alpha(n)))/D \end{bmatrix} + \boldsymbol{\omega}(n) \tag{22}$$

where $T$ is the sampling period, $D$ is the vehicle length and $\boldsymbol{\omega}(n)$ is white measurement noise. The sampling period is chosen as $T = 5ms$ and the vehicle length is assumed to be $D = 300cm$ [32].

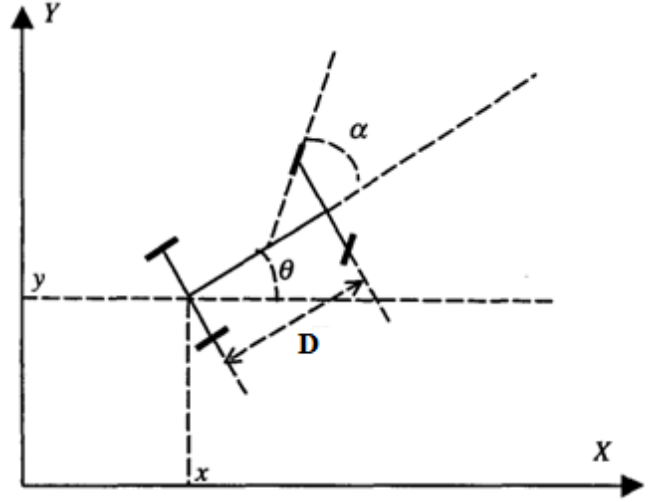

Fig. 4. Autonomous vehicle [32].

To demonstrate the performance of model predictive control using wavelet neural network with feedforward component, we apply our networked control methodology to the autonomous vehicle. We present two simulations scenarios. In the first scenario, the delay from sensor to model and delay form controller to actuator are fixed. This corresponds to the case of a private network for the control system. In the second scenario, network delay is random with a triangular probability distribution function. In both scenarios, measurement noise is assumed to be Gaussian white noise with a variance of 0.1. In both scenarios delayed measurements are discarded.

In both scenarios, all the network parameters are initialized with random values from a normal distribution with variance 0.5 and the WNN is not pretuned. The input to the wavelet neural network is $\mathbf{u}_N(k) = [v(k), \alpha(k), x(k), y(k), \theta(k)]^T$ and the target output of the wavelet neural network is $[x(k+1), y(k+1), \theta(k+1)]^T$. The optimum number of hidden layer nodes was found to be $m = 5$. The learning rate of the network parameters is assumed to be $\gamma = 0.1$. In the controller cost function, the weight of the control inputs is $\rho_1 = \rho_2 = 0.1$ and the weight of the prediction errors is $\xi_j = 0.01, j \in \{1,\ldots,J\}$. A tradeoff between control input magnitude, tracking performance and computational burden is needed to choose appropriate values for $N_p$ and $N_u$. In this simulations, $N_p = N_u = 15$ was found to yield satisfactory performance. The autonomous vehicle is assumed to have a length $D = 300cm$ and the sampling period is chosen as $T_s = 5ms$.

### A. Fixed network delay

In the first scenario, both sensor to controller and controller to actuator network delays are assumed to be $0.1s$. This is a large delay for a sampling period of $T_s = 5ms$ where the MPC receives the measurement and the actuator receives the control action after a delay of 20 sampling periods. To mitigate the effect of this large network delay, the controller predicts 55 samples and uses the last $N_p = 15$ samples to calculate the control action. This provides good compensation for the network delays, assuming that the prediction accuracy is satisfactory and allows the autonomous vehicle to follow the desired trajectory.

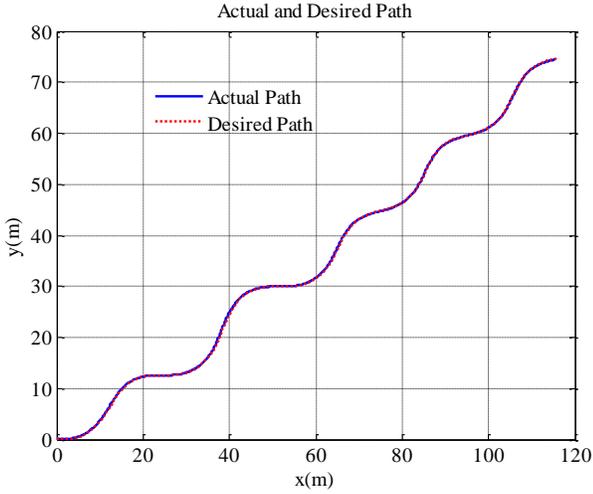

Fig. 5. Tracking of a curved line

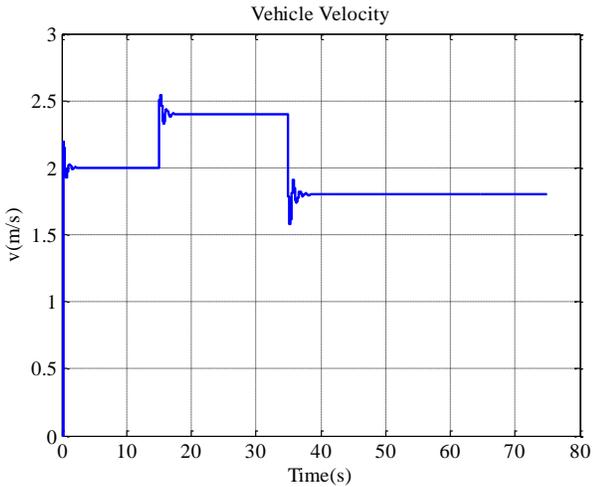

Fig. 6. Vehicle Velocity Calculated by Model Predictive controller

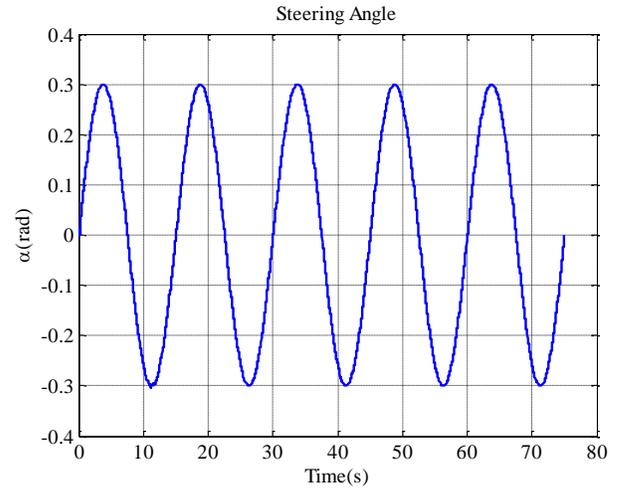

Fig. 7. Steering angle Calculated by Model Predictive controller

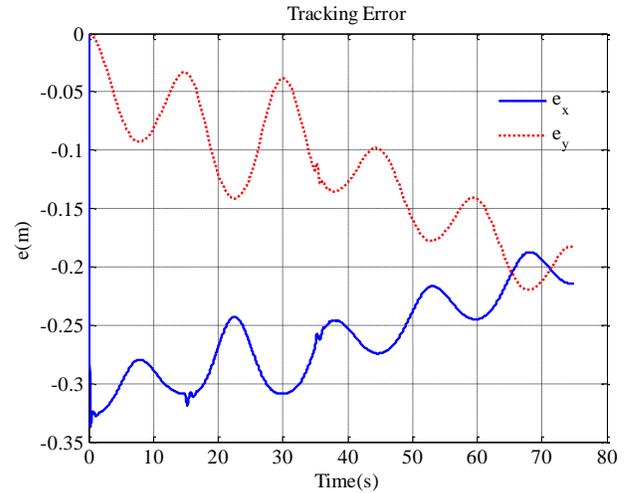

Fig. 8. Tracking error of curved path

Fig. 5. shows a desired curved path together with the vehicle path as it tracks it. Fig. 6 and Fig. 7 show the control inputs for the vehicle. The vehicle must increase and decrease its speed at the appropriate time to be able to track the desired path. Fig.6 shows that using extended prediction and control horizons, controller is able to increase and decrease the velocity at the right time to minimize the tracking error. Fig. 8 shows the tracking error of the curved path. At time $t = 0$, because there is no pretuning on the network parameters, the tracking error is large, but as time progresses the controller identifies the model of the system and reduces the tracking error. The tracking error after 70 seconds is about $20 cm$ in the $x$ and $y$ directions, which is small in comparison to vehicle length of $D = 300 cm$. The tracking error decreases slowly due to large network delay and the online identification process but it asymptotically approaches zero after the shown simulation period.

### B. Random network delay

In the second scenario, network delay is random with triangular probability density function. Thus, only the upper and lower bound of the network delay are known. The packet loss occurs in the network. Packet loss and network delay can be combined and considered as a single delay both for sensor to MPC or for controller to actuator. Therefore, network parameters are updated every $\tau$ seconds with $\tau$ having a triangular probability density function. The sum of network delay and packet loss follows a triangular probability density function with lower limit of $a = 0.005$, upper limit $b = 0.1$, and mode $c = 0.05$. For a sampling period of $T_s = 5ms$, the minimum delay both from sensor to MPC and from MPC to actuator is at least one sampling period. To compensate for the network delay and packet loss in the network, the controller predicts 35 samples beyond the current time, which is equal to twice the mean network delay. This improves the tracking performance of the controller.

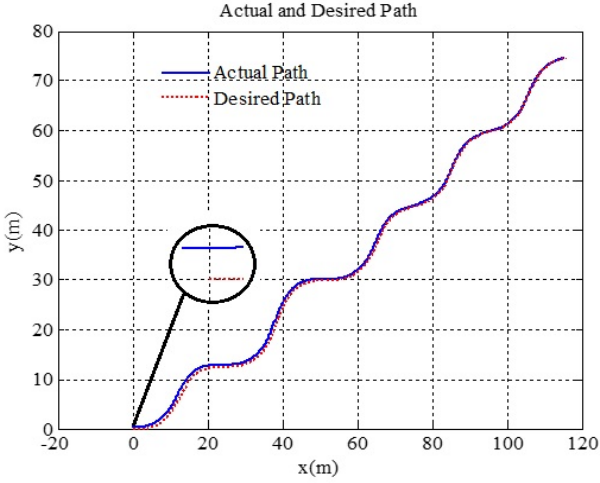

Fig. 9. Tracking of a curved line

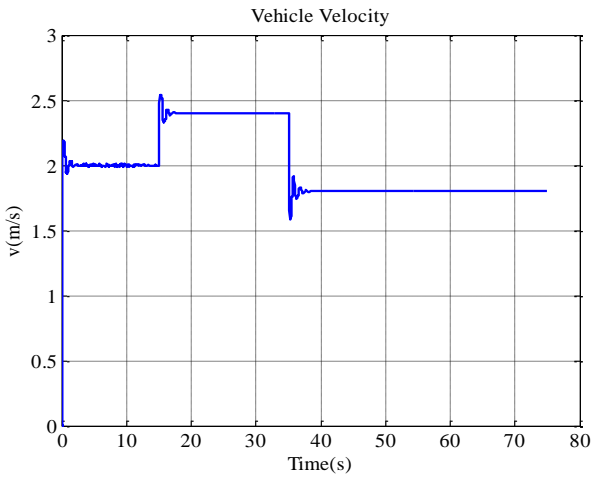

Fig. 10. Vehicle velocity

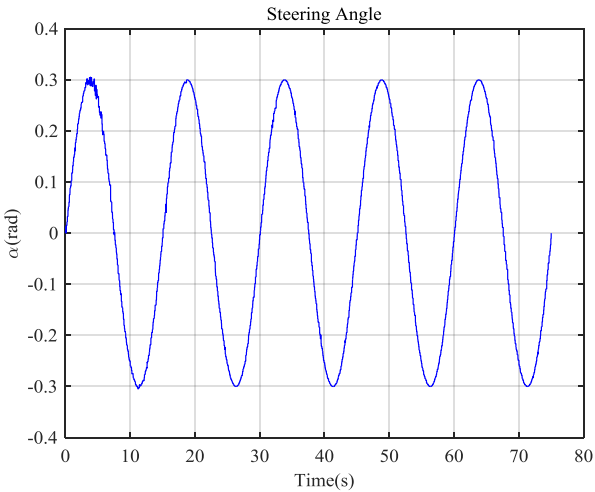

Fig. 11. Steering angle

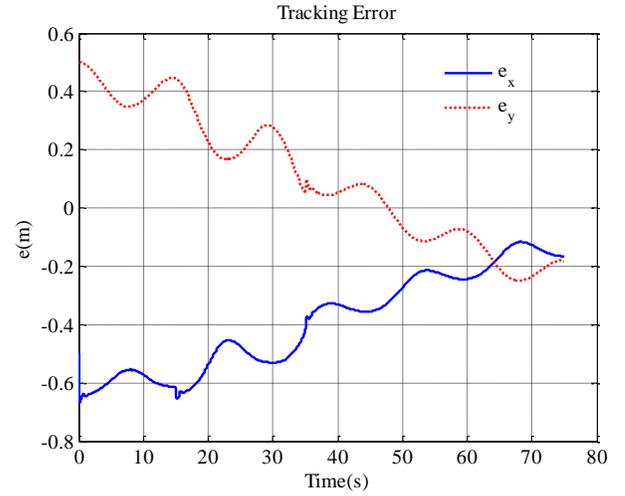

Fig. 12. Tracking error for random delay scenario

Fig. 9 shows the tracking result for random network delay. The vehicle is not on the desired path at the beginning of the simulation but the controller can successfully guide the vehicle toward desired path and track it. Since the vehicle is not on the desired track and the WNN was not pretuned, there is a big tracking error at $t = 0$. However, the network quickly learns the behavior of vehicle and the controller reduces the error. Fig. 10 and Fig. 11 show the control input calculated by the MPC. As in the first scenario, the controller produces smooth control action to track the desired path. The vehicle velocity is dependent on the desired path which may require it to speed up in some parts and slow down in other parts as shown in Fig. 10. Fig. 11 shows the smooth steering angle calculated by the controller. Fig. 12 shows the tracking error for random network delay scenario. The final tracking error in the $x$ direction is about $e_x = 16cm$ and final tracking error in $y$ direction is $e_y = 18cm$. Considering the vehicle length of $D = 300cm$ and large network delay and packet loss, the error is acceptable.

Table I. shows the mean square identification error of $x$, $y$ and $\theta$. The very small mean square identification error shows that the wavelet neural network with feedforward component can efficiently identify the model of autonomous vehicle. In the random network delay scenario, packet loss and the initial position error of the vehicle increase the mean square identification error. Nevertheless, the error is still very small and the network effectively identifies the model of the system.

Table I. Mean square identification error

|  | $x$ | $y$ | $\theta$ |
|---|---|---|---|
| Fixed delay | 0.0363 | 0.0281 | 0.0152 |
| Random delay | 0.0411 | 0.0332 | 0.0211 |

## V. Conclusion

In this study we used model predictive controller along with wavelet neural network with feedforward component to online identification and control of nonlinear system. The feedforward component reduces the number of hidden layer nodes and accelerates the learning and therefore, makes the model more suitable for online identification and control applications. Model predictive controller uses the wavelet neural network with feedforward component to predict the future outputs of the plant over extended prediction horizon. By optimization of controller cost function over extended prediction horizon, controller finds the future control inputs. Simulation results show that this methodology can compensate the effect of fixed and random network delay and packet loss in the network and provide a satisfactory tracking performance. The Lyapunov theory is used to prove the stability of the model predictive controller. Future work will be application of the methodology to unmanned aerial vehicle.

## VI. APPENDIX I

To simplify the stability analysis of the controller, we assume that all $\xi_j = \xi$ and all $\rho_i = \rho$ to obtain

$$I_{N_u} + \xi_i \rho_i H^T = \begin{bmatrix} 1+\xi\rho & \xi\rho & 0 & \dots & 0 \\ 0 & 1+\xi\rho & -\xi\rho & \dots & 0 \\ 0 & 0 & 1+\xi\rho & \dots & 0 \\ 0 & \dots & \ddots & \ddots & -\xi\rho \\ 0 & \dots & 0 & 0 & 1+\xi\rho \end{bmatrix} \quad (22)$$

Assuming $1 + \xi\rho = k_1$ and $\xi\rho = k_2$ we have

$$(I_{N_u} + \xi\rho H^T)^{-1} = F = [f_{i,j}] \quad (23)$$

where

$$f_{i,j} = \begin{cases} \frac{1}{k_1} & , i = j \\ \frac{(k_2/k_1)^{j-i}}{k_1} & , j > i \\ 0 & , j < i \end{cases} \quad (24)$$

Due to upper triangular structure of $F$, we have

$$\lambda_{min}(F) = \cdots = \lambda_{max}(F) = \frac{1}{1+\xi\rho} \quad (25)$$

Rewrite

$$\sum_{j=1}^{J} G_{yj,u_t}^T(n) e_{yj,c}(n)$$

$$= [G_{y1,u_t}^T(n), \dots, G_{yJ,u_t}^T(n)] \begin{bmatrix} e_{y1,c}(n) \\ \vdots \\ e_{yJ,c}(n) \end{bmatrix} \quad (26)$$

Define $G_{Tt}(n) = [G_{y1,u_t}^T(n), \dots, G_{yJ,u_t}^T(n)]$ and $e_T(n) = [e_{y1,c}^T(n), \dots, e_{yJ,c}^T(n)]^T$, then write $\Delta e_{yj,c}(n+1) = e_{yj,c}(n+1) - e_{yj,c}(n)$ as

$$\Delta e_{yj,c}(n+1) = \sum_{t=1}^{n_u} \frac{\partial e_{yj,c}(n+1)}{\partial u_i(n)} \Delta u_i(n)$$

$$= -\sum_{t=1}^{n_u} G_{yj,u_t}(n) R G_{Tt}(n) e_T(n) \quad (27)$$

**Theorem 1:** Assuming that $\|e_{yj,c}\| < k_e, j = 1, \dots, J$, the wavelet based GPC is stable if $\eta < 2(1+\lambda\rho)\Upsilon$ where

$$\Upsilon = \min_{j \in \{1,\dots,J\}} \left\{ \frac{1}{\sum_{i=1}^{n_u} \|G_{yj,u_i}(n)\| \|G_{Ti}(n)\|} \right\} \quad (28)$$

Proof: Consider the discrete Lyapunov function $V(n) = \frac{1}{2}\sum_{j=1}^{J} e_{yj,c}^T(n+1) e_{yj,c}(n+1)$. The change in the Lyapunov function is:

$$\Delta V(n) = V(n+1) - V(n)$$
$$= \frac{1}{2} \sum_{j=1}^{J} \Delta e_{yj,c}^T(n+1)(2e_{yj,c}(n+1) + \Delta e_{yj,c}(n+1)) \quad (29)$$

Define $R = \eta(I_{N_u} + \xi\rho H^T)^{-1}$. $R$ is positive definite and all of its eigenvalues are equal to $\eta/(1+\xi\rho)$. Using (34), $\Delta V(n)$ can be rewritten as

$$\Delta V(n) = \sum_{j=1}^{J} \left[ -\sum_{i=1}^{n_u} e_T^T(n) G_{Ti}^T(n) R^T G_{yj,u_i}^T(n) e_{yj,c}(n) + \frac{1}{2} \left\| \sum_{i=1}^{n_u} e_T^T(n) G_{Ti}^T R^T G_{yj,u_i}^T(n) \right\|^2 \right] \quad (37)$$

$\Delta V(n)$ is the sum of $J$ terms with each term corresponding to one output. For simplicity, we consider the derivatives with respect to each output separately, i.e. we consider the inequality

$$\frac{1}{2} \left\| \sum_{i=1}^{n_u} e_T^T(n) G_{Ti}^T(n) R^T G_{yj,u_i}^T(n) \right\|^2$$
$$< \sum_{i=1}^{n_u} e_T^T(n) G_{Ti}^T(n) R^T G_{yj,u_i}^T(n) e_{yj,c}(n), \quad (39)$$
$$j = 1, \dots, J$$

Using the inequality $\|\omega\| - \|\tau\| \le |\|\omega\| - \|\tau\|| \le \|\omega - \tau\|$ we require

$$\left\| \sum_{i=1}^{n_u} e_T^T(n) G_{Ti}^T(n) R^T G_{yj,u_i}^T(n) \right\| < 2\|e_{yj,c}(n)\|, \quad (40)$$
$$j = 1, \dots, J$$

Since $\|e_{yj,c}(n)\| \le \|e_T(n)\|$, to have negative definite $\Delta V(n)$ we require

$$\eta < 2(1+\lambda\rho)\Upsilon \quad (40)$$

where

$$\Upsilon = \min_{j \in \{1,\dots,J\}} \left\{ \frac{1}{\sum_{i=1}^{n_u} \|G_{yj,u_i}(n)\| \|G_{Ti}(n)\|} \right\} \quad (41)$$